\documentclass[a4paper,twocolumn,11pt, accepted=2026-07-22]{quantumarticle}
\pdfoutput=1
\usepackage[utf8]{inputenc}
\usepackage[english]{babel}
\usepackage[T1]{fontenc}
\usepackage{amsmath}
\usepackage{hyperref}
\usepackage[numbers]{natbib}
\usepackage{tikz}
\usepackage{lipsum}
\usepackage{mathtools}
\usepackage{graphicx}  
\usepackage{bm}
\usepackage{bbold}
\usepackage{physics}
\usepackage{xcolor}
\usepackage{enumitem}
\usepackage{amsmath, amssymb} 
\usepackage{natbib}
\usepackage{physics}

\newcommand{\der}[2]{\partial_{#1} #2}

\newcommand\blfootnote[1]{%
  \begingroup
  \renewcommand\thefootnote{}\footnote{#1}%
  \addtocounter{footnote}{-1}%
  \endgroup
}
\begin{document}
\title{Entanglement growth in the dark intervals of a locally monitored free-fermion chain }

\author{Giovanni Di Fresco*}

\affiliation{Dipartimento di Fisica e Chimica “Emilio Segrè", Group of Interdisciplinary Theoretical Physics, Università degli studi di
Palermo, Viale delle Scienze, Ed. 18, I-90128 Palermo, Italy}

\author{Youenn Le Gal* }

\affiliation{JEIP, UAR 3573 CNRS, Coll\`ege de France, PSL Research University, 75321 Paris Cedex 05, France}

\author{Davide Valenti}
\affiliation{Dipartimento di Fisica e Chimica “Emilio Segrè", Group of Interdisciplinary Theoretical Physics, Università degli studi di
Palermo, Viale delle Scienze, Ed. 18, I-90128 Palermo, Italy}

\author{Marco Schir\`o}
\affiliation{JEIP, UAR 3573 CNRS, Coll\`ege de France, PSL Research University, 75321 Paris Cedex 05, France}
 
\author{Angelo Carollo }
\affiliation{Dipartimento di Fisica e Chimica “Emilio Segrè", Group of Interdisciplinary Theoretical Physics, Università degli studi di
Palermo, Viale delle Scienze, Ed. 18, I-90128 Palermo, Italy}

\blfootnote{* These two authors contributed equally to the work}

\maketitle
\begin{abstract}
We consider a free fermionic chain with monitoring of the particle density on a single site of the chain and study the entanglement dynamics of quantum jump trajectories. We show that the entanglement entropy grows in time towards a stationary state which display volume law scaling of the entropy, in stark contrast with both the unitary dynamics after a local quench and the no-click limit corresponding to full post-selection. We explain the extensive entanglement growth as a consequence of the peculiar distribution of quantum jumps in time, which display superpoissonian waiting time distribution characterised by a bunching of quantum jumps followed by long dark intervals where no-clicks are detected, akin to the distribution of fluorescence light in a driven atom. We show that the presence of dark intervals is the key feature to explain the effect and that by increasing the number of sites which are monitored the volume law scaling gives way to the Zeno effect and its associated area law.
\end{abstract}

\section{Introduction}
The fluorescence light emitted from a driven atom is characterised by periods of darkness interrupted by random, abrupt periods of brightness, corresponding to quantum jumps between atomic states~\cite{Cohen-Tannoudji_1986,cook1985possibility}. These early studies in atomic physics led to the first observations of quantum jumps~\cite{bergquist1986observation,nagourney1986shelved}, which have become since then an experimental reality, observed in a variety of platforms in atomic physics, quantum optics and solid-state physics~\cite{minev2019tocatch}. Quantum jumps also motivated the foundation of the modern theory of quantum trajectories~\cite{ueda1990,dalibard1992wavefunction,gardiner1992wave}, describing the physics of a quantum system which is continuously monitored by an external environment~\cite{plenio1998quantum,wiseman2009quantummeasurementand}.
\newline
\indent In recent years the effect of quantum jumps and quantum measurements have attracted fresh new interest in the context of open quantum many-body systems~\cite{daley2014quantum,fazio2024manybodyopenquantumsystems} and their measurement-induced phase transitions~\cite{nahum2019measurementandentanglement,li2018quantumzenoeffect,li2019measurementdrivenentanglement}. Here the focus is on the entanglement properties of quantum many-body trajectories and how these are affected by the competition between unitary evolution and measurements~\cite{legal2024entanglementdynamicsmonitoredsystems}. 
\newline
\indent The standard setting where this transition has been studied involves the interplay between unitary evolution, described by a many-body Hamiltonian or by a random circuit, and a finite density of local projective measurements~\cite{fisher2023randomquantumcircuits}. Alternatively, weak-measurement protocols  have been also discussed, such as in the quantum jump or quantum state diffusion unravelling of the Lindblad evolution. In these cases one assume every site of the lattice to be exposed to a monitoring stochastic protocol. In all these examples the effect of measurements result in an entanglement transition from a volume-law scaling~\cite{fuji2020measurementinducedquantum,lunt2020measurementinducedentanglement,dogger2022generalizedquantummeasurements,xing2023interactions,altland2022dynamics}, or sub-volume in the case of non-interacting systems~\cite{cao2019entanglementina,fidkowski2021howdynamicalquantum,coppola2022growthofentanglement,loio2023purificationtimescalesin,poboiko2023theoryoffree,jian2023measurementinducedentanglement,fava2023nonlinearsigmamodels,carisch2023quantifying,jin2023measurementinduced,alberton2021entanglementtransitionin,vanregemortel2021entanglement,turkeshi2021measurementinducedentanglement,botzung2021engineereddissipationinduced,bao2021symmetryenrichedphases,turkeshi2022entanglementtransitionsfrom,piccitto2022entanglementtransitionsin,kells2023topological,paviglianiti2023multipartite,muller2022measurementinduced,soares2024entanglementtransitionparticlelosses}, to an area law where the system is frozen due to the quantum Zeno effect. This elusive criticality is hidden in the fluctuations of the monitoring protocol, while it remains transparent to the averaged state described by a Lindblad master equation. Experimental investigations of measurement-induced transitions have appeared~\cite{noel2021measurementinducedquantum,koh2022experimentalrealizationof,hoke2023quantuminformationphases} but are plagued by the postselection problem, i.e. the exponential overhead cost in sampling a quantum trajectory associated to a given measurement outcome, even though solutions in certain fine-tuned cases have been proposed~\cite{ippoliti2021postselectionfreeentanglement,passarelli2023postselectionfree,garratt2024probing}.

In this work, we explore the role of monitored dynamics and quantum jumps in a minimal model, where monitoring only occurs locally on a given site of an otherwise free fermionic chain. Local perturbations on quantum many-body systems, i.e. quantum impurities, have been long-known to induce non-trivial effects~\cite{anderson1967infrared,bettelheim2006orthogonality,schiro2014transient,Krapivsky_2019,tonielli2019orthogonality,froml2019fluctuation,pocklington2022stabilizing}. Here we show that a non-unitary local monitoring results in a striking growth of entanglement entropy, which slowly saturates to an extensive value scaling with the volume of the subsystem size. This is in stark contrast with both the unitary case and the fully postselected, no-click limit. Therefore our results show that quantum jumps are crucial in generating entanglement. We explain this result by looking at the statistics of quantum jumps, which we reveal to display periods of bunching, interspersed with long dark intervals, similar to the resonance fluorescence of a driven atom. We demonstrate that the presence of these dark periods is the key to extensive entanglement generation and that by monitoring more sites the entanglement generally decreases, recovering the area-law regime expected when each lattice site is monitored.

\begin{figure*}[!t] 
\centering 
\includegraphics[width=0.99\textwidth]{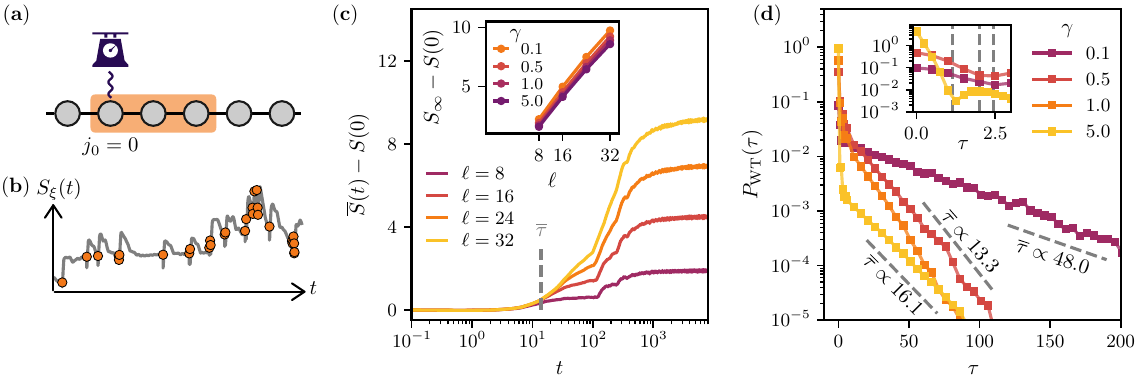} 
    \caption{Entanglement dynamics for a locally monitored fermionic chain. (a) Cartoon of the tight-binding chain where the site $j_0$ is continuously monitored. The yellow region depicts the partition $A$ of size $\ell$ that we use to compute the bi-partite von Neumann entanglement entropy. (b) Time evolution of the entanglement entropy evaluated along a typical trajectory. (c) Dynamics of the averaged entanglement entropy for a measurement rate $\gamma=0.5$ and different values of the subsystem size $\ell$. In the inset is shown the linear scaling of the long time limit of the entanglement entropy with subsystem sizes $\ell$ for different measurement rates $\gamma$. (d) Waiting time distribution of the jumps for different measurement rates $\gamma$. The short-time behavior of the distribution is shown in the inset, to highlight the first exponential decay. Parameters: $J = 0.5,\, j_0 = 0,\,  L = 128,\, \nu = 1/4$}
    \label{fig:oms_ent}
\end{figure*} 

\section{Model and Monitoring Protocol} 
We consider a lattice model of non-interacting $1D$ fermions, where a continuous monitoring procedure at site $j_0$ is performed. The model can be described by the Hamiltonian 
\begin{align}
H_0 = -J\sum_j c^\dagger_{j+1} c_{j} + h.c.    
\end{align}
with $L$ sites and periodic boundary conditions, $J$ being the hopping strength. We consider a monitoring quantum jump protocol where the system wave function evolves randomly in time according to the stochastic Schrödinger equation~\cite{dalibard1992wavefunction,ueda1990,gardiner1992wave,wiseman2009quantummeasurementand}
\begin{align}
         \der{t}{\ket{\psi}} &= -idt\left\{H_0 - \frac{i}{2}\left(L^{\dagger}_{j_0}L_{j_0} - \langle L^{\dagger}_{j_0}L_{j_0}\rangle\right)\right\}\ket{\psi}\nonumber\\
         &+ d\xi\left\{ \frac{L_{j_0}}{\sqrt{\langle L^{\dagger}_{j_0}L_{j_0} \rangle}} - 1\right\}\ket{\psi}  \label{eq:sto_schrodi}
\end{align}
where $d\xi(t) \in \{0,1\}$ is an inhomogeneous Poisson increment such that $P(d\xi(t) = 1 ) = dt \bra{\psi(t)}  L_{j_0}^\dagger L_{j_0}\ket{\psi(t)} $. Monitoring is implemented via a single jump operator of the form $L_{j_0} = \sqrt{\gamma}n_{j_0}$ (as depicted in the sketch of Fig.~\ref{fig:oms_ent} (a) ). 

A generic quantum trajectory described in Eq.~(\ref{eq:sto_schrodi}) results in a series of evolutions under an effective non-Hermitian Hamiltonian
$H_{NH} = H_0 - i\frac{\gamma}{2}n_{j_0}$, (corresponding to the first line of Eq.~(\ref{eq:sto_schrodi})), interrupted by random quantum jumps where the number of fermions at site $j_0$ is measured. Two limits of this problem have been studied: (i) the fully post-selected evolution corresponding to a trajectory with no quantum jumps, also called the no-click limit, where the problem reduces to free fermions in an imaginary scattering potential~\cite{stefanini2023orthogonalitycatastropheluttingerliquid,dolgirev2020nongaussian} and (ii) the dynamics of the average state described by a Lindblad master equation for a free fermion chain subject to a local dephasing~\cite{dolgirev2020nongaussian}.

In this work we are interested in studying the entanglement structure of quantum trajectories evolving under the protocol described above. We consider a pure initial state, a Fermi sea with constant filling, $\nu$. Such a choice makes the no-click evolution a local quench, since it is an eigenstate of the Hermitian part of $H_{NH}$. Different choices of initial state do not change qualitatively the results, as we discuss in Appendix A.
Since the system remains Gaussian along the quantum trajectory, Eq.~(\ref{eq:sto_schrodi}) can be efficiently simulated~\cite{legal2024entanglementdynamicsmonitoredsystems}.
Furthermore, since the purity is conserved by the stochastic dynamics we can characterize the entanglement via the von Neumann entanglement entropy defined as $S_\xi (t) = - \mathrm{Tr}_A \left[ \rho_\xi^A(t) \ln \rho_\xi^A (t) \right] $ where we have introduced a bi-partition A,B (cf Fig.~\ref{fig:oms_ent} (a) ), with the reduced density matrix $\rho^A_\xi (t) = \mathrm{Tr}_B \lvert \psi_\xi (t) \rangle \langle \psi_\xi (t) \rvert $. In the following, we will discuss in particular the behavior of the entanglement entropy averaged over the measurement noise as
\begin{equation}
    \overline{S}(t) = \int \mathcal{D} \xi P(\xi) S_\xi (t)\,.
    \label{eq:av_entanglement}
\end{equation}
 \section{Entanglement properties} \label{sec:ent_discussion}
We start by discussing the dynamics of entanglement entropy, plotted in Fig.~\ref{fig:oms_ent}(c) for $\gamma=0.5$ and different subsystem sizes $\ell$. In order to highlight the growth due to the monitoring we subtract the value of the entropy at time $t=0$, corresponding to the critical Fermi sea. We begin by observing a remarkable production of entanglement and a threshold behavior. After some initial transient regime, where the entanglement entropy remains close to the initial value, we see a sizeable growth with time and a strong dependence on subsystem size. Furthermore, the entanglement entropy reaches a steady state on a notably slow timescale. While this timescale for free fermions is usually proportional to the sub-system length, here we find it to be much longer and compatible with an exponential scaling (see Appendix B).

Another remarkable feature is that the steady-state value of the entanglement entropy grows linearly with the subsystem size $\ell$, as illustrated in Fig.~\ref{fig:oms_ent} (c), i.e. it displays volume-law scaling. Unlike typical free-fermion MIPT~\cite{cao2019entanglementina,alberton2021entanglementtransitionin,coppola2022growthofentanglement}, where the entire chain is monitored, we do not observe an area law scaling regime, even at reasonably large measurement rates $\gamma$. Crucially, this is a feature of the local monitoring protocol: indeed, if we increase the number of sites which are monitored we see that the entanglement entropy decreases and the volume law scaling turns into an area-law one (see Appendix G). The dramatic effect of a single monitoring protocol on the entanglement dynamics is one of the major results of this work. These results are particularly striking when compared to the no-click limit evolution, corresponding to a local quench by an imaginary scattering potential, but also if compared to the analogous unitary dynamics, i.e. when the quench is realized by a real scattering potential. In both cases indeed the entanglement entropy  produced due to the evolution would quickly saturate to a value independent on sub-system size (see Appendix A and D).

This highlights the fact that quantum jumps in our problem play a key role for the entanglement generation. To further understand this point, in the following we will take a closer look to the statistics of quantum jumps.

\section{Statistics of Jumps Waiting times }
Looking at a characteristic single shot evolution of the entanglement entropy, plotted in Fig.~\ref{fig:oms_ent} (b), already suggests something peculiar occurs in this problem. The way in which quantum jumps are distributed displays two alternating behaviors, characterized by bunching periods and dark periods, respectively. Specifically, the dynamics appears to consist of series of quantum jumps (plotted as orange dots)  occurring in close succession (bunching periods), interspersed with long intervals during which the system does not jump, and the dynamics is governed by non-Hermitian evolution (dark periods). This intuition is confirmed by the waiting time distribution (WTD) of quantum jumps, displayed in Fig.~\ref{fig:oms_ent} (d), which exhibits two distinct timescales, thereby showing a significant deviation from the typical Poissonian distribution seen in jumps unraveling~\footnote{The statistics of jumps do not always follow a Poissonian distribution throughout the entire dynamics; however, it usually is reached in the steady state. This is not the case here}. Most quantum jumps occur at very short waiting time $\tau$, following a Poissonian distribution. However, we also observe a sizable amount of quantum jumps occurring at a much longer waiting times, giving rise to a statistically significant tails in the distribution. This kind of distribution is known as super-Poissonian, and it often arise in contexts where events are bunched.  We will explain how this distinctive feature plays a key role in generating the observed entanglement dynamics.

\section{Entanglement growth, dark periods and bath disentangling}
The emergence of these two time scales in the WTD enables the presence of dark intervals in the monitoring dynamics. The jumps that occur during the bunching periods contribute to the initial exponential decay in the WTD (see ~Fig.\ref{fig:oms_ent} (d)), while the jumps following the dark periods are responsible for the long tail. An interesting aspect to discern is whether the increase in entanglement can be associated either to the presence of bunching or to the long periods during which the measurement device does not click. 
\begin{figure*}[!t] 
    \centering
\includegraphics[width=0.99\textwidth]{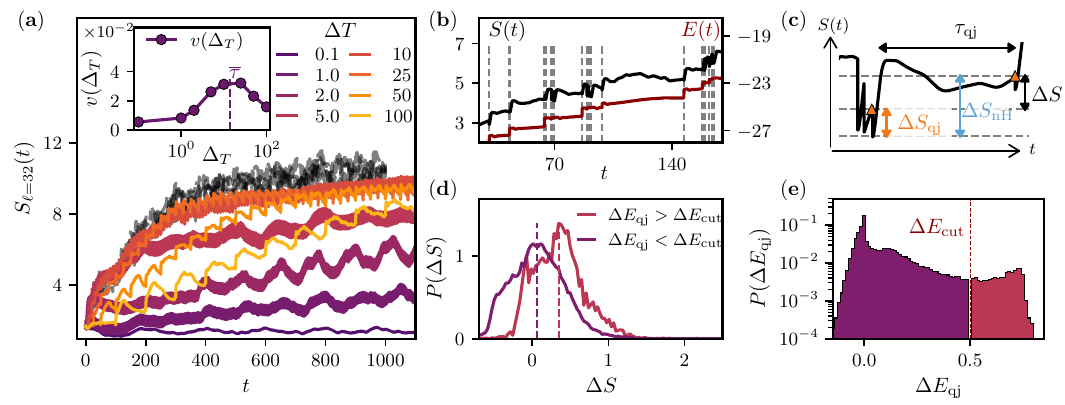}
    \caption{
    Entanglement growth mechanism.  (a) Entanglement dynamics of quantum trajectories trajectories where the waiting time is fixed to $\Delta T$. Dark curves are  typical random trajectories and the inset shows rate of entanglement growth $v(\Delta T)$ where $\overline{\tau}$ is the average waiting time  of the long tail of the full WTD. 
  (b) Dynamics of the tight binding energy $E(t)$ (red curve) and of the entanglement entropy $S(t)$ (black curve). (c) Sketch defining the different quantities, thereby $\Delta S_\mathrm{qj}$ denotes the change of entanglement during one QJ, and $\tau_{qj}$ is the waiting time following a jump.  (d) Marginal distribution $P(\Delta S)$ of the change of entanglement $\Delta S=\Delta S_\mathrm{qj} + \Delta S_\mathrm{nH}$  conditioned on the energy variation being larger or smaller $\Delta E_\mathrm{cut}$ 
     (e) Distribution $P(\Delta E_\mathrm{qj})$ of the change of energy during a QJ. 
     Parameters: $\gamma=0.5,\, J = 0.5,\, j_0 = 0,\,  L = 128,\, \nu = 1/4$.}
    \label{fig:ent_to_ener}
\end{figure*}
We now argue that the key to the extensive volume-law like entanglement growth in our problem of local monitoring consist in these dark periods during which the evolution is driven by a non-Hermitian Hamiltonian, starting however from a different initial state after each random quantum jump. To investigate this, in Fig.~\ref{fig:ent_to_ener} (a) we analyze trajectories with fixed waiting times $\Delta T$, and we evaluate the entanglement dynamics along them. For this purpose, we generate trajectories in which the jump times are fixed independently of the corresponding jump probabilities. In principle, the same ensemble of trajectories could be obtained by post-selection from the stochastic dynamics. However, such a procedure would be numerically impractical, since the probability of selecting trajectories with prescribed jump times becomes rapidly suppressed. We emphasize that, in the limit of sufficient sampling, the two procedures yield the same result. The rate of entanglement growth $v(\Delta T)$ ( defined as the slope of the entanglement growth before saturation ), displays a clear non-monotonic dependence on the fixed waiting time $\Delta T$.

Importantly, we see that $v(\Delta T)$ reaches its optimal value when $\Delta T$  matches the average waiting time $\overline{\tau}$ observed during the dark intervals (cf. Fig.~\ref{fig:oms_ent} (d)).
This analysis shows that the essential factor for achieving significant entanglement growth is the presence of these dark periods. To gain further insight into why long dark periods enable entanglement growth we now look in more details at the statistics of the entanglement entropy changes along a quantum trajectory. This metric was recently introduced to understand the mutual role of non-Hermitian evolution and quantum jumps for entanglement phase transitions~\cite{legal2024entanglementdynamicsmonitoredsystems}. As we show in Appendix D, when applied to our model this analysis reveals that the net effect of quantum jump events is to decrease the entanglement entropy, while the non-Hermitian evolution is responsible for its growth. Interestingly the entanglement loss due to a jump depends non-monotonously on the length of the waiting time preceding the event, suggesting that bunched jumps have less effect than the one happening after a dark period of order $\overline{\tau}$. 

Although detrimental for entanglement, quantum jumps play a key role in resetting the state onto which the non-Hermitian evolution will act on, allowing extensive entanglement to be generated. This resetting turns out to be substantial, even though the jump operator acts only on a single site of the system - a kind of orthogonality catastrophe. To understand this point we look at the dynamics of the free fermion energy $E(t)=\expval{H_0}{\psi_\xi(t)}$ along a quantum trajectory and compare it to the evolution of the entanglement entropy, see Fig.~\ref{fig:ent_to_ener} (b). As shown in Fig.~\ref{fig:ent_to_ener} (e), when sampling the energy variation, $\Delta E_\mathrm{qj}$ across an event we observe a bimodal distribution, one mode for large energy variations and one mode for small ones. We separate these two behaviors by introducing $\Delta E_\mathrm{cut}$ as the point where the distribution displays a minimum between the modes. 

These energy variations are correlated to the entanglement changes, large ones occurring usually after a long dark period and come with a large change in the entanglement entropy variation between the event and the following one, $\Delta S = \Delta S_\mathrm{qj} + \Delta S_\mathrm{nH}$( these quantities are defined in Fig.~\ref{fig:ent_to_ener}). Indeed, if we now sample the statistics of entanglement entropy changes conditioned on the amount of energy variation, we see two different behaviors (Fig.~\ref{fig:ent_to_ener}~(d)). Most of weak energy variations typically do not affect entanglement entropy, while the large ones induce a finite net increase, corresponding to the histogram $P_{\Delta E_\mathrm{qj}>\Delta E_\mathrm{cut}}(\Delta S)$ peaked around a finite non-zero value.

We can now understand at least qualitatively the origin of this extensive entanglement production: when a quantum jump hits the monitored site after a long dark period of no-clicks it induces a non-trivial effect across the chain, disentangling part of it and increasing its energy, thus bringing the system far away from equilibrium and leading to an increase in entanglement entropy. This is further supported by the analysis of the correlation between energy and entanglement entropy changes after a quantum jump, as well as of density-density correlations along a given quantum trajectory showing clear bath-disentangling around the monitored site (see Appendix E). As a result, the non-Hermitian evolution which follows this type of events can be considered akin to a global quantum quench, where part of the chain is perturbed and re-entangled rapidly and efficiently, rather than a local one where entanglement entropy growth can be, at best, logarithmic. Out of this balance between loss due to quantum jumps and gain due to entangling evolution, a steady state is eventually reached after a certain number of dark periods. In the no-click evolution, on the other hand, although the dynamics is still driven by the same non-Hermitian Hamiltonian, the initial state is an eigenstate of the free-fermion Hamiltonian. This implies that the perturbation to the state is only due to the measurement back-action (responsible for the non-Hermiticity), whose effect is too local to induce any sizable change in the bulk of the chain. As such it is not surprising that the entanglement production in the no-click limit remains small (see Appendix C). This also clarifies why the entanglement production is non-monotonous with $\Delta T$: if we increase $\Delta T$ too much, the system remains stuck with the non-Hermitian Hamiltonian in this steady state for too long, slowing down the entanglement production. Our results therefore showcase how local measurements can have dramatic effect on many-body states. Importantly, while our discussion has been focused on the quantum jump protocol, the main effect survives also in the presence of projective measurements (see Appendix F).

\section{Conclusions} 
In this work we have discussed the effect of local on-site monitoring via quantum jumps on a fermionic chain. Remarkably, we have shown that such a local non-unitary perturbation results in a slow growth of entanglement entropy which saturates to a volume law value, in striking contrast to both the unitary case and the no-click limit where only subextensive entanglement can be generated by local perturbations. We have highlighted the key role of quantum jumps and their super-Poissonian WTD in explaining the entanglement production. In particular we have shown that, while quantum jumps decrease the entanglement entropy they also reset the many-body state disentangling the monitored site and its surrounding from the rest of the chain and thus leaving space for the non-Hermitian evolution to produce entanglement during the long dark intervals. Our results represent one of the simplest yet nontrivial effects due to monitoring, which crucially comes with low overhead in postselection, given the local monitoring and the slow dynamics leading to few quantum jumps. As such, our results could be experimentally verified in quantum simulators with on-site addressing, such as Rydberg arrays~\cite{nguyen2018towards,ravon2023array,browaeys2020many}.
We expect that the effect of interactions along the chain would not change significantly the entanglement pattern as long as the disentangling effect of quantum measurements remain effective. A detailed study of the interplay between local measurements and interactions represents an exciting perspective for future studies.
 
\begin{acknowledgments}
Y.L.G. acknowledges useful discussions with M. Vanhoecke. Authors acknowledge computational resources on the Collège de France IPH cluster. AC acknowledges support from European Union – Next Generation EU through projects: Eurostart 2022 Topological atom-photon interactions for quantum technologies MUR D.M. 737/2021. AC and DV acknowledge PRIN 2022- PNRR no. P202253RLY Harnessing topological phases for quantum technologies; THENCE – Partenariato Esteso NQSTI – PE00000023 – Spoke 2. MS acknowledges funding from the European Research Council (ERC) under the European Union’s Horizon 2020 research and innovation programme (Grant agreement No. 101002955 — CONQUER). G.D.F. acknowledges support from the PNRR MUR project PE0000023–NQSTI.
\end{acknowledgments}
\section*{Author contributions}
GDF and YLG developed the code and performed all the numerical and analytical analyses. MS and AC coordinated the work. All authors contributed to the discussion of the results and to the writing of the manuscript. The authors used generative AI tools only to proofread the manuscript.

\appendix
\onecolumn
\section{Role of Initial Condition and Global vs Local Quenches}
\begin{figure*}[t!]
	\centering\includegraphics[width=0.8\textwidth]{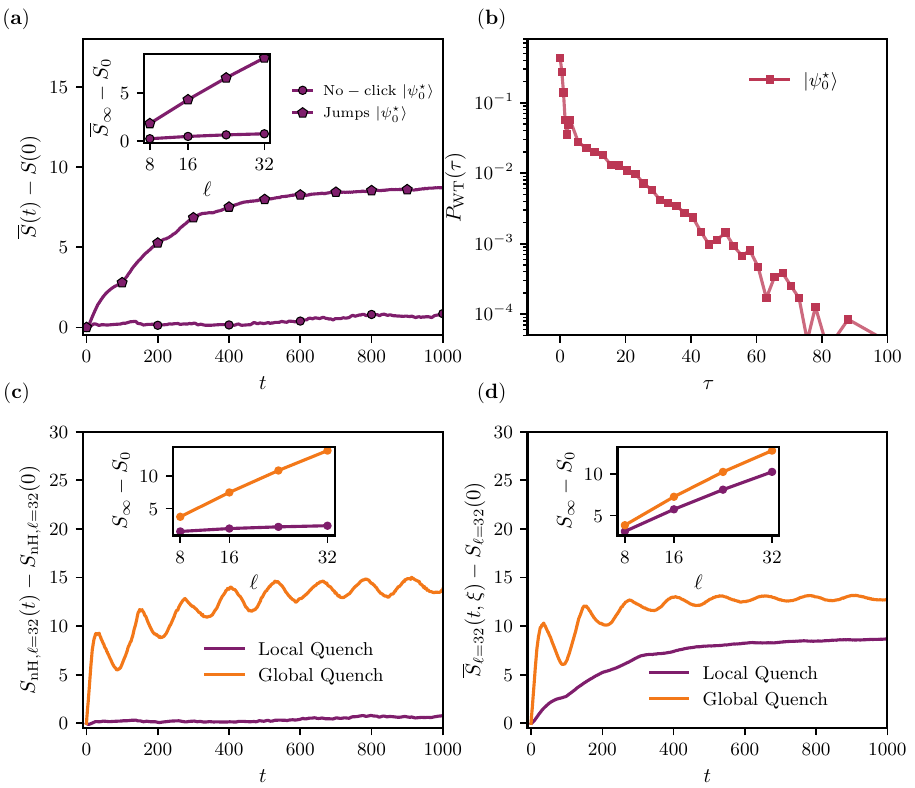}
	\caption{a) Entanglement dynamics (subsystem of size $\ell=32$) for the no-click evolution (circles) and averaged entanglement dynamics over the full sample of trajectories under monitored evolution (pentagons); in both cases the initial state differs from the one in the main text: $\lvert \psi_0^{\star} \rangle$ is the ground state of a tight binding Hamiltonian with the local chemical potential $V = \gamma_0 n_{j_0}/2$ for $\gamma_0 =0.4$.  b) WTD for the monitored dynamics starting with the initial state $\lvert \psi_0^{\star} \rangle$. c) Entanglement dynamics in the no-click limit for two different quenches: when the initial state is a product state (global quench, orange) and when the initial state is the ground state of a tight binding Hamiltonian, as in the main text (local quench, purple). d)  Averaged entanglement dynamics over the full sample of trajectories under monitored dynamics with the same initial states as in panel c). Other Parameters : $L=128$, $\gamma=0.5$  } 
    \label{fig:glo_vs_loc}
\end{figure*}
In this section we discuss the role of the initial state for the entanglement entropy dynamics and scaling in the steady state. In the main text, we have discussed the case in which the initial state is the ground state of the tight-binding chain, i.e. a Fermi sea with a given filling.  We now consider two different scenarios: (i) a local quench and (ii) a global quench. In the first case, the system is prepared initially in the ground-state of the tight-binding chain with an extra chemical potential localized at the monitored site. The value of the chemical potential is then suddenly changed at time $t>0$.
This implements a local unitary quench in addition to the sudden switching of the monitoring protocol. As is well documented, this local unitary quench by itself creates excitations, but not enough to generate volume-law entanglement~\cite{stephan2011local}. In Fig.~\ref{fig:glo_vs_loc}(a) we present the entanglement entropy dynamics in the no-click limit, which now corresponds to a local quench of a complex-valued chemical potential, and the full quantum jump dynamics. We consider as initial state $\lvert \psi_0^{\star} \rangle$ the ground state of a tight binding chain with the real potential $V = \gamma_0 n_{j_0}/2$ where $\gamma_0=0.4$. Remarkably the volume law growth (see inset) is observed only in presence of quantum jumps, this further confirms that our main result is not dependent on the specific choice of the initial state, but rather a genuine steady-state feature. Furthermore, in Fig.~\ref{fig:glo_vs_loc}(b) we show that the initial state is not changing the waiting-time distribution structure that we discuss in the main text, namely a heavy tail corresponding to long dark intervals with no jumps.

The second scenario is discussed in Fig.~\ref{fig:glo_vs_loc}c) and d) and it is compared to the local quench case.  Our global quench is implemented by considering for initial state a trivial product state in real-space and the dynamics generated by the tight-binding chain is thus a global quench of the hopping, in addition to the local monitoring. Not surprisingly, in this case the steady-state entanglement entropy always shows volume-law scaling, which is essentially driven by the unitary dynamics. This is confirmed in Fig.~\ref{fig:glo_vs_loc}(c) where we consider the no-click limit of this protocol. As we see from comparison with Fig.~\ref{fig:glo_vs_loc}(d), showing the average entanglement entropy under jump dynamics. Under this protocol, there is essentially no difference between no-click and jump dynamics - all interesting effects are masked by the global quantum quench. For comparison, we plot in the same panels the case discussed in the main text, corresponding to a local quench. Here the difference between no-click dynamics and full quantum jumps is striking. \ref{fig:glo_vs_loc}(d) summarize in a nutshell the main finding of our work: a local quantum jump monitoring gives rise to a volume-law entanglement growth in the steady-state, which is comparable to the one generated by a global quantum quench.

\section{Saturation Time Scale for Entanglement Entropy}

 \begin{figure*}[t!] 
\centering\includegraphics[width=0.7\textwidth]{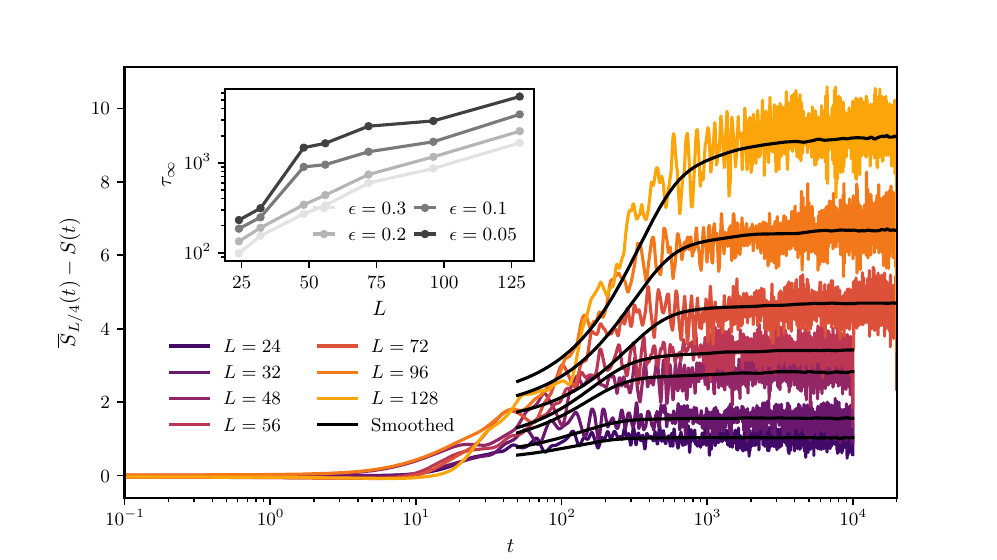} 
    \caption{Entanglement entropy dynamics  for chains of varying lengths $L$. The subsystem is for all curves of size $L/4$. The black solid line corresponds to the smoothened dynamics used to determine the saturation time $\tau_\infty$. The inset shows the saturation time $\tau_\infty$ scaling with system size $L$ for different $\epsilon$. Parameters: $J = 0.5,\, \gamma = 0.5,\, j_0 = 0,\,  \nu = 0.25$. } 
    \label{fig:steady-state-time}
\end{figure*}
In this section, we analyze the time scale required for the entanglement entropy of our system to reach a steady state, focusing on how the saturation time depends on system size. This process is very slow, which is consistent with the local nature of the monitoring. Getting this saturation time $\tau_\infty$ is thus challenging numerically due to the long simulation time required to reach the steady state and also to the sizable number of trajectories needed to reduce sampling noise and thus probe the saturation of entanglement. To this purpose, we compute the entanglement entropy of a system of size $L$, considering $L/4$ of the chain as a subsystem. By varying the total length, we can observe the relationship between saturation time and system size. 

In the main plot of Fig.~\ref{fig:steady-state-time}, we show the entanglement entropy dynamics $S_{L/4}(t)$ for chains of different lengths. To attempt to evaluate $\tau_\infty(L)$ we use a Gaussian convolution filter to smoothen the entanglement dynamics. Then we obtain the saturation time as earliest $\tau_\infty$ such that $|S(t) -S_\infty |\leq \epsilon$ for all $t>\tau_\infty$ ($S_\infty$ is determined by doing the time integrated average of the late time of the simulation). For each system size the black solid line corresponds to the smoothened entanglement dynamics. 

The inset shows that the scaling is compatible with an exponential growth of the time $\tau_\infty$ to reach the steady state with system size $L$. We note that, while our numerical analysis does not show a quantitative convergence with $\varepsilon$, the qualitative scaling of $\tau_{\infty}$ with $L$ is compatible with an exponential for every $\varepsilon$. The fact that the time scale for the relaxation of the entanglement is so long, as compared, say, to the unitary dynamics under a Hermitian scattering where the scaling is ballistic, is remarkable. We note that the scaling of the relaxation time with $L$ and the volume-law scaling of the steady-state entanglement appear also to be compatible with the slow dynamics of the entanglement entropy reported in the main text.

\section{Entanglement Dynamics in the No-Click limit}
\label{app:no-clcick}
\begin{figure*}[h]
	\centering\includegraphics[width=0.8\textwidth]{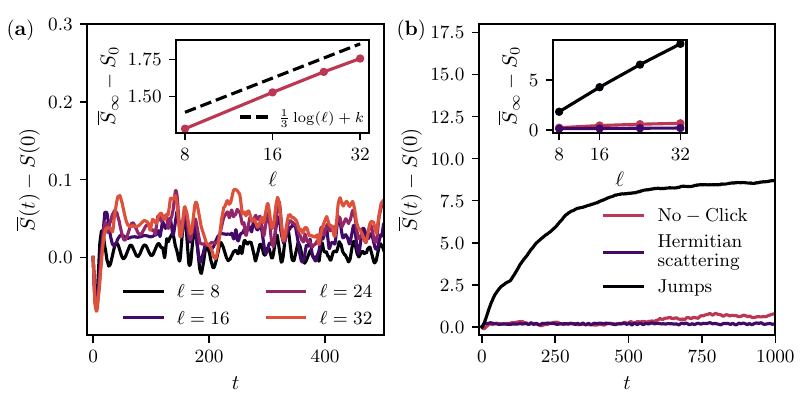}
	\caption{Panel (a): Entanglement entropy as a function of time for different cuts 
    $\ell$. The inset shows the scaling of the entanglement with respect to system size. The red solid line represents the numerical data, while the dashed black line indicates the theoretical prediction for the ground state of the free-fermion model. Panel (b): Comparison between the entanglement growth of the no-click scenario (red line) and the averaged entanglement entropy over the full unraveling (black line), as well as the hermitian evolution with a real scattering potential (purple line). Parameters: $J = 0.5,\, \gamma = 0.3,\, j_0 = 0,\, L = 128,\, \nu = 0.25$. }
	\label{fig:nh}
\end{figure*}

In this section, we examine the entanglement behavior in one specific trajectory of the full jump unraveling: the so-called no-click limit. This is an exponentially rare, deterministic scenario where no clicks occur throughout the evolution. Although the no-click properties may not capture the broader behavior of the full unraveling, they often reveal intriguing aspects that are worth exploring. The no-click evolution is governed by a non-Hermitian Hamiltonian, and understanding its characteristics is valuable for studying the dynamics of the no-click contribution to each trajectory, and hence, of the full  system’s dynamics. Furthermore, since our primary focus is on entanglement dynamics, gaining insight into what occurs in the absence of measurement events is essential to gauge the relative performance of click and no-click evolutions.
\newline
\indent In this section, we want to emphasize a point already highlighted in the main text: the presence of quantum jumps is the key ingredient that allows our system to generate significant entanglement. To demonstrate this point in Fig.~\ref{fig:nh}(a) we show the evolution of the entanglement entropy for the no-click Hamiltonian defined in Eq.(3) of the main text, for three subsystem sizes $\ell$. We see that the entanglement growth is very limited and displays a weak subsystem size dependence that quickly saturates. Note that in this plot we subtract the value at time $t=0$ to focus on the contribution coming from the time evolution and not from the initial state, a critical Fermi sea with logarithmic scaling of the entanglement entropy which dominates the overall scaling (see inset). 
\newline
In panel (b) of Fig.~\ref{fig:nh}, we compare the entanglement entropy of the quantum jump trajectories for the same parameters, showing the mean entanglement entropy over the jump trajectories alongside the no-click trajectory. As it appears clearly, the curve averaged over the jump trajectories displays significantly higher entanglement content. In panel (b), for the sake of completeness, we also compare these two scenarios with the entanglement that would be generated by a sudden quench of a Hermitian scattering potential $V=\frac{\gamma}{2}n_{j_0}$ (as opposed to the non-unitary scattering $- i\frac{\gamma}{2}n_{j_0}$ of the no-click Hamiltonian in Eq.~3). This comparison shows that, in this scenario as well, the production of entanglement is negligible.
\newline
\indent There are two additional aspects of the non-Hermitian Hamiltonian that we will discuss without explicitly presenting the results. First, it is known that this Hamiltonian has an exceptional point at $\gamma/2J =1$~\cite{stefanini2023orthogonalitycatastropheluttingerliquid}. Interestingly, the entanglement structure, as measured by entanglement entropy, does not seem to be affected by this point, either in no-click trajectories or in the full unraveling (see panel (d) of Fig.~1). The second aspect concerns the role of the impurity position. While its position doesn’t alter the qualitative behavior of the system, it does slow down the dynamics.
\section{Statistics of Entanglement Gain and Loss}
\begin{figure*}[h] 
\includegraphics[width=0.99\textwidth]{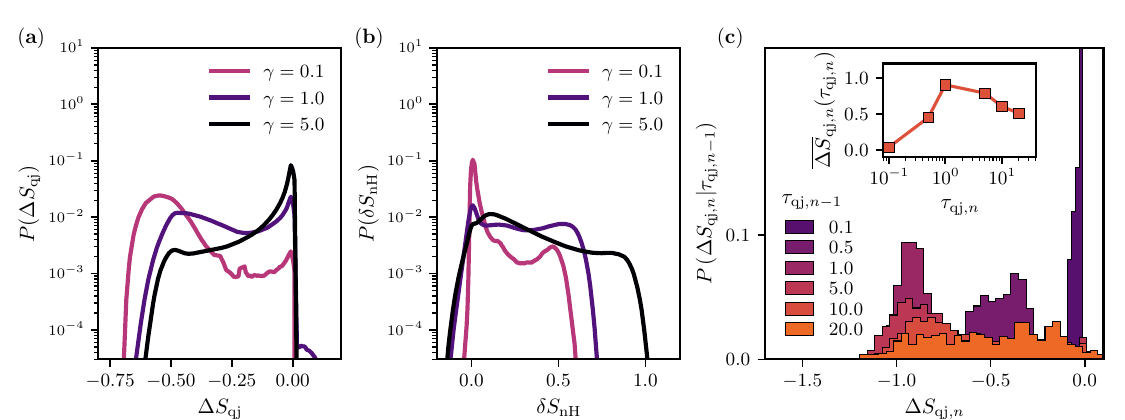}
\caption{Quantum Jumps Statistics. (a) Statistics of entanglement change due to QJs, $P(
\Delta S_\mathrm{qj})$. See sketch of Fig.~3 (c) for the definition of the quantities.  (b) The corresponding quantity for the non-Hermitian evolution. Here the variation of the entanglement due to the non-Hermitian evolution is renormalized by the waiting time $\tau$ such that $\delta_\mathrm{nH} = \Delta_\mathrm{nH}/\tau$ as done in \cite{legal2024entanglementdynamicsmonitoredsystems} (c) The statistics of entanglement change due to QJs, $P(\Delta S_{\mathrm{qj},n}\lvert \tau_{\mathrm{qj},n-1} )$ , conditioned on the waiting time $\tau_{\mathrm{qj},n-1}$ just before the jump. In the inset the averaged statistics $\overline{\Delta S}_{\mathrm{qj},n}(\tau_{\mathrm{qj},n-1} ) = \int_{\Delta S \mathrm{qj}} \Delta S_{\mathrm{qj},n} P( \Delta S_{\mathrm{qj},n}\lvert \tau_{\mathrm{qj},n-1})$ is plotted. Parameters: $\gamma=0.5,\, J = 0.5,\, j_0 = 0,\,  L = 128,\, \nu = 1/4$. }
    \label{fig:post_select_traj2}
\end{figure*}

To understand better the role of quantum jumps and non-Hermitian evolution in the dynamics of our problem we follow Ref.~\cite{legal2024entanglementdynamicsmonitoredsystems} and study the statistics of entanglement gain and loss along a quantum trajectory. The key idea of this approach is that along a quantum jump trajectory there are two types of evolution, a deterministic one driven by the non-Hermitian Hamiltonian followed by a random quantum jump. Here we monitor the change in entanglement entropy after a quantum jump and in-between them, i.e. during the non-Hermitian evolution, and collect two hystograms $P(\Delta S_\mathrm{qj})$ and $P(\delta S_\mathrm{nH})$ that we plot in Fig.~\ref{fig:post_select_traj2}(a-b). A remarkable feature of these distributions is the sharpness of their support: quantum jumps appear to always decrease the entropy while non-Hermitian evolution always to increase it. This property is in stark contrast with the case of globally monitored free fermions, where as observed previously~\cite{legal2024entanglementdynamicsmonitoredsystems} these distributions have tails, implying that a quantum jump has a finite (although small) probability of increase the entanglement entropy.

Moreover, contrary to Ref.~\cite{legal2024entanglementdynamicsmonitoredsystems} in this case if we conditioned the QJ distribution on the preceding waiting time, we observe a significant variation of the statistics of entanglement gain due to quantum jumps (See Fig.~\ref{fig:post_select_traj2}(c) inset). In particular, the shorter the waiting time is the less effect the jump has on the entanglement (which is consistent with a Zeno effect). When the waiting time increases, the effect on the entanglement also increases until it reaches a maximum . For a longer waiting time the distribution spreads and in average the effect on the entanglement of a QJ diminishes. We retrieve a similar situation to Fig.~2(a) with a non-monotonous behavior which complements the analysis of the main text. 

\newpage
\section{Additional Numerical Results}

\begin{figure*}[t!] 
    \centering
    \includegraphics[width=0.90\textwidth]{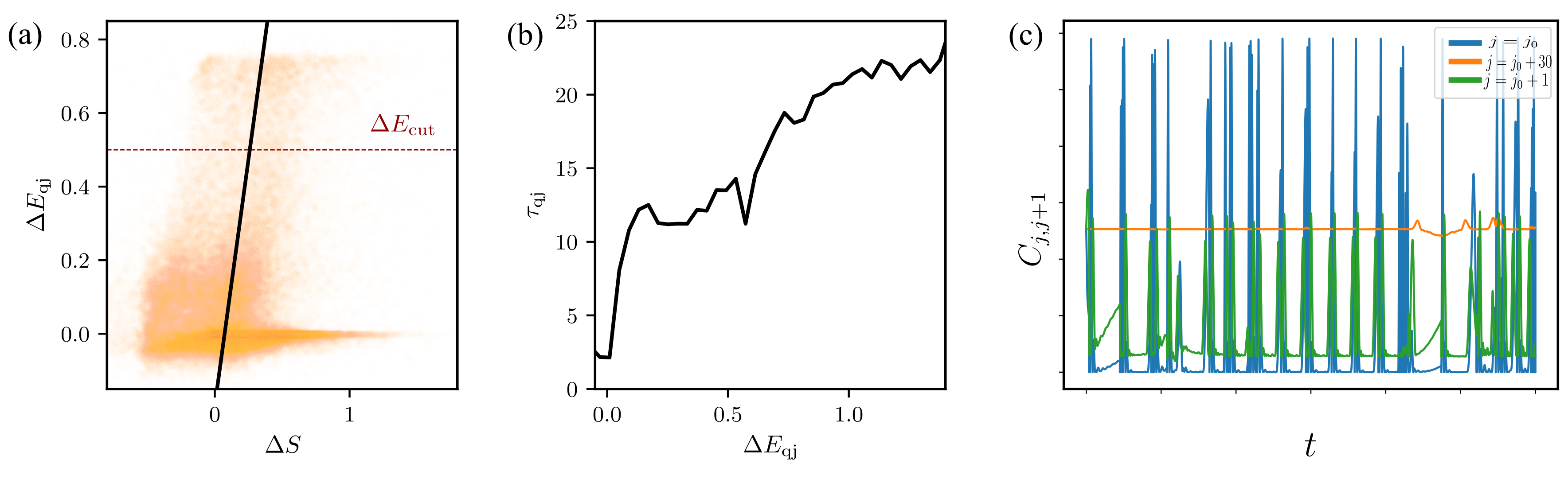} 
    \caption{Additional analysis about the free fermion energy - (a)
    Correlation between the variation of energy during a quantum jump $\Delta E_\mathrm{qj}$ and the change of entanglement $\Delta S=\Delta S_\mathrm{qj} + \Delta S_\mathrm{nH}$. (b) Waiting time as a function of the variation of energy $\Delta E_{\mathrm{qj}}$ occurring in the jump following the waiting period $\tau_{\mathrm{qj}}$. (c) Connected correlation function $C_{j,j+1} = \langle n_j n_{j+1} \rangle - \langle n_j \rangle \langle n_{j+1} \rangle$ for a typical trajectory. The blue curve represents the connected correlation between the monitored site $j_0$ and its neighbor. The green curve shows the connected correlation between two sites in the vicinity of the monitored site. The orange curve corresponds to the connected correlation deep in the bulk.  Parameters: $\gamma=0.5,\, J = 0.5,\, j_0 = 0,\,  L = 128,\, \nu = 1/4$. } 
     \label{fig:corr_s_e}
\end{figure*}

In this section, we provide additional numerical results on the quantum jump dynamics of the free fermion energy $E(t)=\expval{H_0}{\psi_\xi(t)}$, discussed in Fig.~2 of the main text. 
In particular, in Fig.~(\ref{fig:corr_s_e})(b) we plot the waiting time between quantum jumps $\tau_{\mathrm{qj}}$ as a function of the energy change  $\Delta E_{\mathrm{qj}}$ occurring in the jump following the wait. We see that the waiting time grows rapidly with $\Delta E_{\mathrm{qj}}$ suggesting that large energy changes in the free fermion chain occur after large waiting times, corresponding to long dark periods of no-clicks. \\
In Fig.~(\ref{fig:corr_s_e})(a) we plot the correlation between the variation of energy during a quantum jump $\Delta E_\mathrm{qj}$ and the total change to the entanglement entropy $\Delta S=\Delta S_\mathrm{qj} + \Delta S_\mathrm{nH}$, happening between the same jumps and the following one (in this expression $\Delta S_\mathrm{nH}$ is the contribution to the entanglement produced in the non-Hermitian evolution between these two jumps). From this plot we see that there is indeed a correlation between large energy changes in the chain and large change in the entanglement entropy variation between a quantum jump event and the following one. This supports the picture of local quantum jumps having a strong effect on the rest of the fermionic chain, disentangling the bath and increasing its energy bringing it out of equilibrium.

To further support this picture, in Fig.~(\ref{fig:corr_s_e})(c) we plot the connected density-density correlations along a given quantum trajectory, $C_{j,j+1} = \langle n_i n_{j+1} \rangle - \langle n_j \rangle \langle n_{j+1} \rangle$, for different sites $j$.
After each jump, the connected correlations between the monitored site and all other sites vanish (blue curve in the plot). We also show that sites in the vicinity of the monitored one become disentangled after the jump. In particular, if $j_0$ denotes the monitored site, we display in the figure the connected density-density correlation between sites $j_0+1$ and $j_0+2$ (green curve). This demonstrates that quantum jumps decorrelate not only the monitored site but also its neighboring region.

\section{Comparison with Projective Measurements}

\begin{figure*}[h!] 
\centering\includegraphics[width=0.75\textwidth]{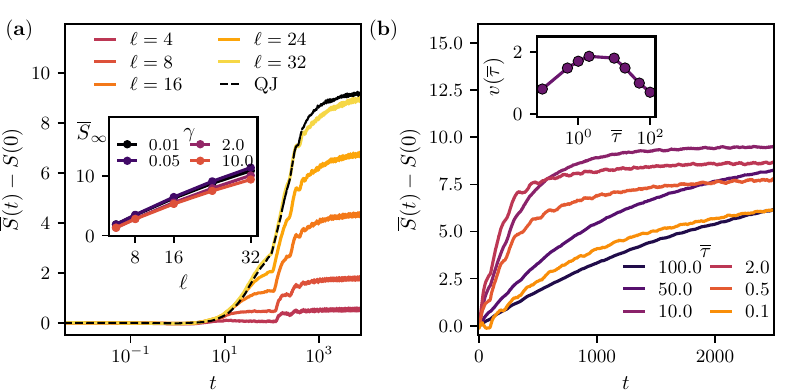} 
    \caption{Entanglement Entropy Dynamics under Projective Measurements. (a) Entanglement dynamics for different values of the subsystem size $\ell$ and $\gamma=0.5$. The black curves shows the entanglement dynamics in the case of QJs for $\gamma=0.5$ and $\ell=32$. In the inset is plotted the long time limit of the entanglement $\overline{S}_\infty$ as a function of the subsystem size $\ell$. (b) Entanglement dynamics of $\ell =32$ and different values of $\gamma = 1/\overline{\tau}$. In the inset we show the growth velocity. Parameters: $J = 0.5,\, j_0 = 0,\, L=128\, \nu = 0.25$.  } 
     \label{fig:unraveling}
\end{figure*}

In this section we compare the results for the entanglement entropy obtained in the main text under quantum jumps unraveling with the dynamics under projective measurements. In particular we consider two projectors  the site $j_0$ given by $n_{j_0}$ and $1 - n_{j_0}$ and a unitary evolution with the free-fermion Hamiltonian within the measurement events. The waiting time between the measurement is chosen to follow a Poissonian law of rate $\gamma$. In Fig.~\ref{fig:unraveling}(a) we plot the average entanglement entropy versus time, for different subsystem sizes $\ell$, and we find qualitatively a very similar behavior to the results of main text (see dashed line for quantum jumps results). In particular the scaling of the entanglement in the steady state is still linear, compatible with a volume law (see inset). In Fig.~\ref{fig:unraveling}(b) we plot the short-time dynamics of the entanglement entropy as a function of the average time between measurement $\bar{\tau}$, which in this case is simply given by $1/\gamma$. Interestingly, we see that the slope of the entanglement entropy $v$ depends non monotonically on the average time (see inset), again in agreement with the general picture in the quantum jump case. We stress that in this case changing $\gamma$ only changes the event distribution in time, thus this strengthens the idea that in the QJs case the non-unitary part is not really important: what really matters is having jumps (here projective measurements) sufficiently spaced in time.  

\section{Monitoring a fraction of sites }

\begin{figure*}[h!] 
	\centering\includegraphics[width=0.85\textwidth]{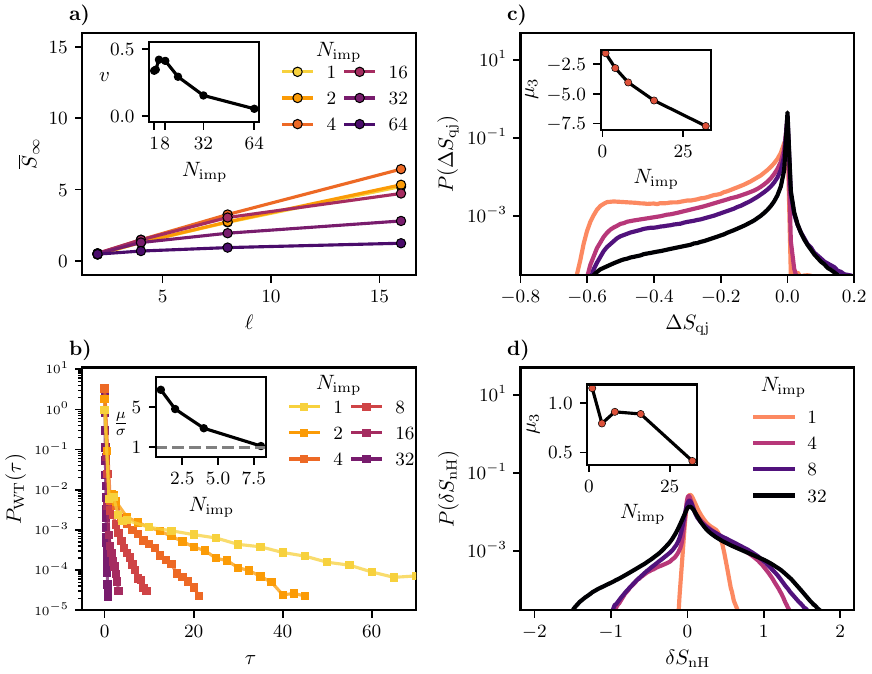} 
	\caption{ The transition in the statistical properties of the model occurs as the number of monitored sites increases. (a) Entanglement entropy scaling for different number of monitoring sites $N_\mathrm{imp}$. The inset shows the coefficient of the linear scaling. (b) Waiting time distribution for different number of monitoring sites $N_\mathrm{imp}$. (c) and (d) Distributions of entanglement change $P(\Delta S_\mathrm{qj})$ during QJs and non-Hermitian $P(\delta S_\mathrm{nH})$. In the insets is plotted the $3^\mathrm{rd}$ moment of the distribution, connected with the skewness.  } 
	\label{fig:many_sites}
\end{figure*}

In this section, we aim to characterize the transition in the statistical properties of our model from the scenario described in the main text, i.e. a local monitoring, to the standard one discussed in the literature on measurement-induced entanglement transitions, where each site is monitored. To this end, instead of monitoring a single site $j_0$, we choose to monitor $N_\mathrm{imp}$ sites at positions $j_0$, \ldots, $j_{N_\mathrm{imp}}$, gradually increasing the number of impurities until they constitute a fraction of the chain length. The results of this analysis are reported in Fig.~\ref{fig:many_sites}. Panel (a) shows that as we increase the number of monitoring sites, the entanglement entropy transitions from volume scaling to an area law, according to results of~\cite{poboiko2023theoryoffree}. Interestingly, this transition from volume to area law is reflected in a change in the waiting time distribution of jumps. Panel (b) shows that, as the number of monitoring sites increases, a transition occurs from super-Poissonian to Poissonian behavior. We can easily see it by looking at the ratio $\mu/\sigma$, which tends to one when increasing the measurement fraction. Additionally, examining the loss and gain distribution of QJs and non-Hermitian evolution, as the number of detectors increases, is instructive (see panels (c) and (d)). In both cases we observe that the asymmetry decreases when $N_\mathrm{imp}$ increases, which means that we lose the sharpness discussed in Fig.~\ref{fig:post_select_traj2}. In the case of QJs big negative variation of the entanglement due to jumps are less likely. 

\newpage

\bibliographystyle{quantum}
\bibliography{main}

\end{document}